\documentclass{emulateapj}
\usepackage{apjfonts}
\usepackage{lscape}

\input{epsf}
\slugcomment{Submitted to the Astronomical Journal}

\shorttitle{Clumping in O-star winds}
\shortauthors{L\'epine \& Moffat}

\begin{document}

\title{Direct spectroscopic observations of clumping in O-star
  winds.}

\author{S\'ebastien L\'epine\altaffilmark{1,2} \& Anthony
  F. J. Moffat\altaffilmark{1,3}}

\altaffiltext{1}{Visiting astronomer, Cerro Tololo Inter-American
  Observatory, National Optical Astronomy Observatories, operated by
  the Association of Universities for Research in Astronomy, Inc.,
  under cooperative agreement with the National Science Foundation.}

\altaffiltext{2}{Department of Astrophysics, Division of Physical Sciences,
American Museum of Natural History, Central Park West at 79th Street,
New York, NY 10024, USA}

\altaffiltext{3}{D\'epartement de Physique, Universit\'e de
  Montr\'eal, Montr\'eal, Qc, Canada, moffat@astro.umontreal.ca}

\begin{abstract}
We report the detection and monitoring of transient substructures in
the radiation-driven winds of five massive, hot stars in different
evolutionary stages. Clumping in the winds of these stars shows up as
variable, narrow subpeaks superposed on their wide, wind-broadened
(optical) emission lines. Similar patterns of emission-line profile
variations are detected in the Of stars $\zeta$ Puppis and HD93129A,
in the more evolved hydrogen-rich, luminous, Of-like WN stars HD93131
and HD93162, and in the more mass-depleted WC star in
$\gamma^2$Velorum. These observations strongly suggest that stochastic
wind clumping is a universal phenomenon in the radiation-driven, hot
winds from all massive stars, with similar clumping factors in all
stages of mass depletion.
\end{abstract}


\section{Introduction}

Evidence has been accumulating in the last two decades that hot
stellar winds are far from being the smooth flows of escaping matter
that were often conveniently assumed. Current observations are of
sufficient quality to show that the strong, hot stellar winds of
bright, population I Wolf-Rayet (WR) stars are systematically pervaded
by inhomogeneities on different scales. Variability on time scales of
minutes to hours is so far best described by a fully clumped wind with
no smooth component \citep{LM99,LEM99}. This clumping may very well
be related to small-scale radiative instabilities leading to strong
shocks in the wind \citep{GO95}.

Is this phenomenon unique to the extremely dense winds of WR stars, or
do other hot stars also have similar globally clumped winds which have
been difficult to detect due to their weaker emission lines?
Subsequent to our WR studies, we found that the Of star $\zeta$ Puppis
shows identical structure in its wind, as found (statistically) in WR
stars \citep{ELM98}: $\zeta$ Pup's optical He II $\lambda$ 4686
emission line varies in exactly the same manner as the isolated HeII
5411 line in WN stars or CIII 5696 in WC stars. The variable emission
component observed at very high S/N in $\zeta$ Pup is also the {\it
  same fraction} of the total emission-line flux as in WR spectra,
whose absolute amplitude is significantly less in O stars. This may
explain why it had remained virtually unnoticed before in O stars. If
this one O star is typical, then we expect to find the same stochastic
emission-line variability in other O-star winds.

Other types of hot stars also exhibit clumpy structure in their winds.
Examples include the [WC] nuclei of planetary nuclei \citep{GMA03},
novae \citep{LSLZ99}, supernovae \citep{M00} and possibly
others that have not been examined properly yet.

Turbulent clumping in hot-star winds has a number of very important
consequences, e.g. among the most important are: (1)  Most current
estimates of mass-loss rates, which are sensitive to the square of the
density, have to be revised downwards {\em by about a factor three}
with important implications for stellar evolution \citep{MR94}. (2)
Since clumps follow the mean flow, on average, they are turning out to
be extremely useful in constraining the empirical wind velocity law,
v(r), that is quite fundamental, but poorly known at present in
hot-star winds; in particular, the convenient, popular interpolation
$\beta$-law appears to be only a rough approximation \citep{LM99}.
(3) Wind clumps may ultimately provide the necessary means of
compression and shielding required to form dust in some WR and nova
winds \citep{Wetal87,CT95,B95,MM07}. This might be relevant for dust
formation in any type of star. The presumed associated decrease in the
mass-loss rates could have a major effect on their evolution, slowing
down the shedding of their envelope and delaying their eventual
transfer to a WR stage. 

\begin{deluxetable*}{llcccll}
\tabletypesize{\scriptsize}
\tablecolumns{7} 
\tablecaption{Stars observed}
 \tablehead{HD number & Other name & RA(2000) & Dec(2000) & V magnitude & Spectral type & Comment}
\startdata 
HD 66811  & $\zeta$ Puppis   & 08 03 35.04 & -40 00 11.3 & 2.21 & O4I(n)f & single runaway \\
HD 93129A & \nodata          & 10 43 57.46 & -59 32 51.3 & 6.97 & O3If*   & visual binary\tablenotemark{1} \\
HD 93131  & \nodata          & 10 43 52.25 & -60 07 04.0 & 6.48 & WN6ha   & single \\
HD 93162  & \nodata          & 10 44 10.33 & -59 43 11.4 & 8.11 & WN6ha   & spectroscopic binary\tablenotemark{2} \\
HD 68273  & $\gamma^2$ Velorum & 08 09 31.95 & -47 20 11.7 & 1.81 & WC8+O7.5III-V  & spectroscopic binary\tablenotemark{3} \\
\enddata
\tablenotetext{1}{Companion 0.05\arcsec from the primary \citep{Netal04}.}
\tablenotetext{2}{Orbital period 207.7 days \citep{Getal06}.}
\tablenotetext{3}{Orbital period 78.5 days \citep{DM00}.}
\end{deluxetable*}

One way out, however, is via the still poorly-understood, intermediate,
luminous blue variable (LBV) stage, which may provide the necessary
strong episodic mass-loss to ultimately allow O stars to pass to WR
stars \citep{S07}. In the meantime, massive-star models have been
vastly improved. Including rotation in the models, which is generally
quite rapid for O stars, will tend to feed more H-rich fuel into the
core region and increase the luminosity for a given mass, thereby
allowing more O stars to become WR, and relaxing the requirement for
unexplainably high mass-loss rates in the HRD isochrone fits to
observations \citep{MM00}. Therefore, it appears particularly urgent
to examine the clumping question in O-star winds of all types, so that
evolutionary models for massive stars on/near the main sequence can be
checked/constrained. This will increase our confidence in the age
determinations of clusters and starbursts, which are turning out to be
crucial tracers of star formation in the early Universe. It is also
relevant in the question of the progenitors of the slow gamma-ray
bursts, believed to arise in rapidly rotating WR stars
\citep{H07}. The rotation rate will be higher if the progenitor loses
less mass on the way to becoming a WR star. In this context, it is
interesting to note that clumping is still as strong in SMC WR stars
at Z$_{\odot}/5$ as at Z$_{\odot}$ in the Galaxy \citep{Metal07}.

Very recent studies using the indirect approach have revealed
decreased mass-loss rates in O-star winds by factors of 3 or more
\citep{BLH05} to 10 or more \citep{FMP06}. These are based on weak
resonance and other lines in the far UV with the FUSE satellite. In
the UV, the emission lines are stronger, but often saturated and in
any case, it is difficult to obtain sufficiently high S/N. A recent
complementary approach in the UV has been successful, however, in
revealing variable features in the strong P Cygni absorption edges in
the FUSE UV spectrum of the single WC8 star WR135 by \citet{Metal06},
linking the presence of overdense clumps with observed shocks.

In this study, we look for direct evidence for wind clumping in
optical emission-lines generated in O-star and related winds. Clumping
is detected in the acceleration zone near the base of the wind as
transient, moving subpeaks on the emission-line profiles. Among O
stars, Of stars are the easiest to examine, since they have emission
lines in the optical, where it is possible to obtain high S/N spectra
of these normally weak lines.

\begin{figure*}
\epsscale{1.15}
\plottwo{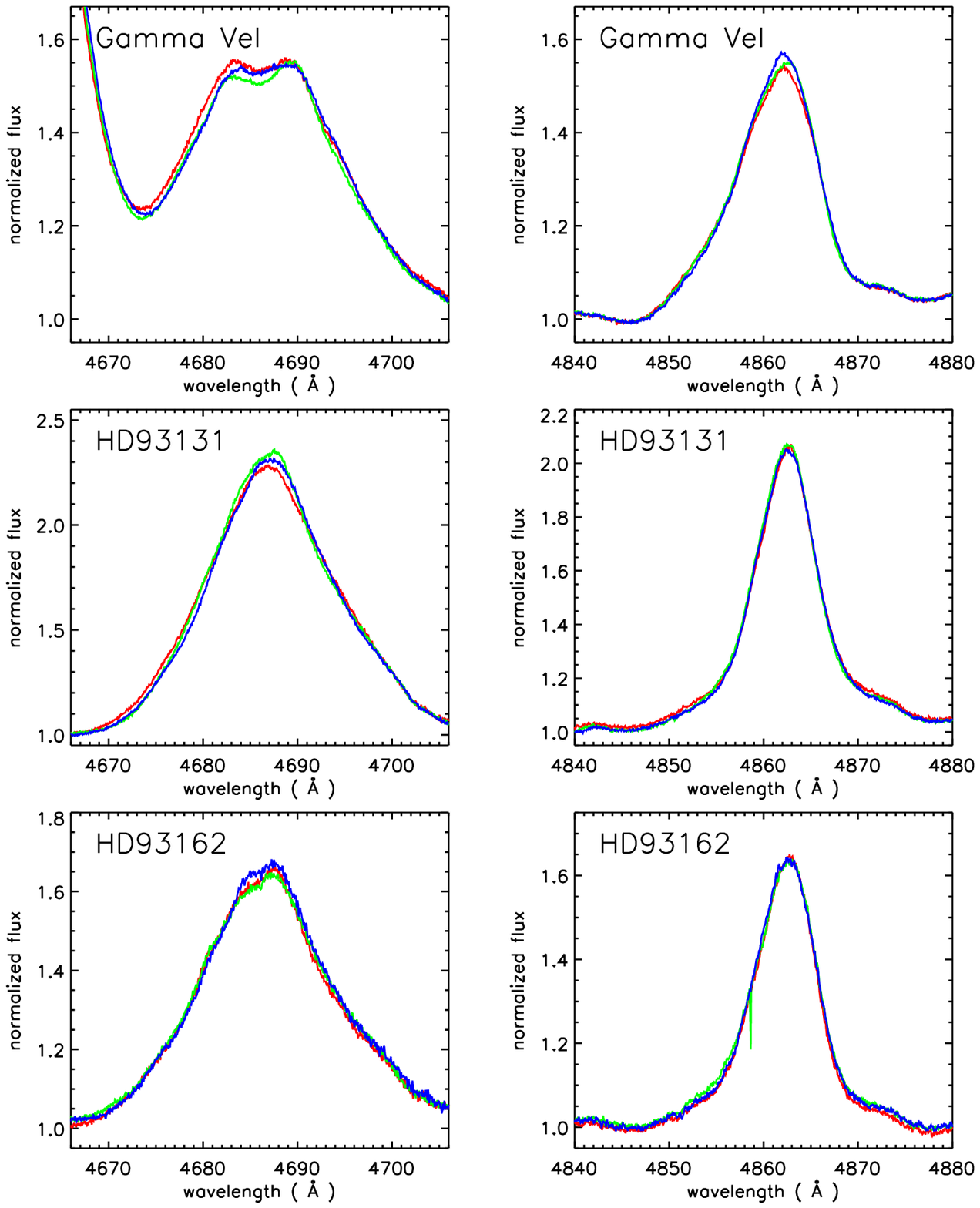}{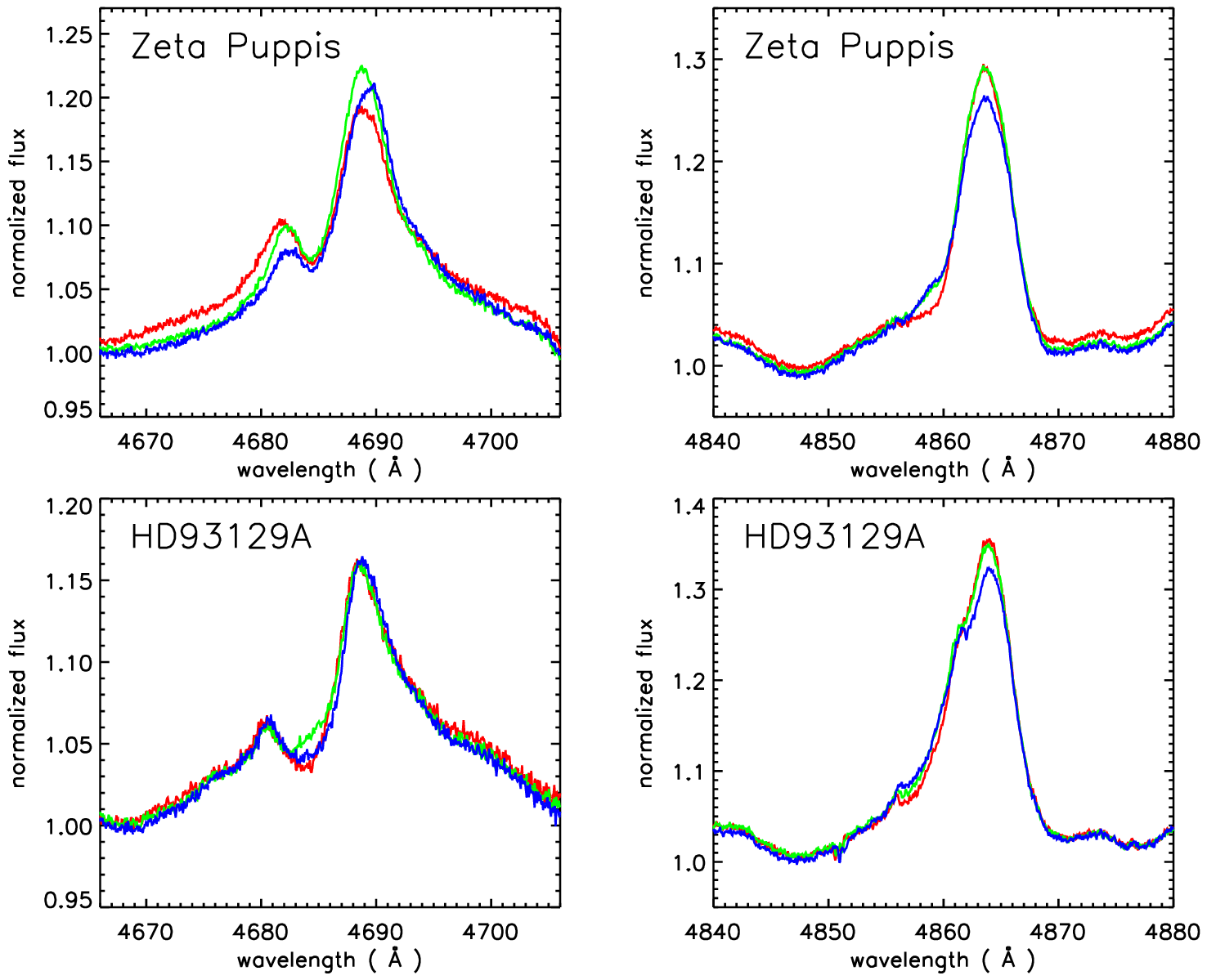}
\caption{Nightly mean spectra of our five program stars, listed in
  Table 1. For each star, two orders from the \'echelle spectrogram
  are shown, with the wavelength regions around the broad, intense
  emission lines of HeII 4686 (left) and HeII/H$\beta$ 4860
  (right). The mean spectra for the nights of 2000 Jan 25, 26 and 27
  are shown in red, green and blue, respectively.}
\end{figure*}

\begin{figure}
\epsscale{1.0}
\plotone{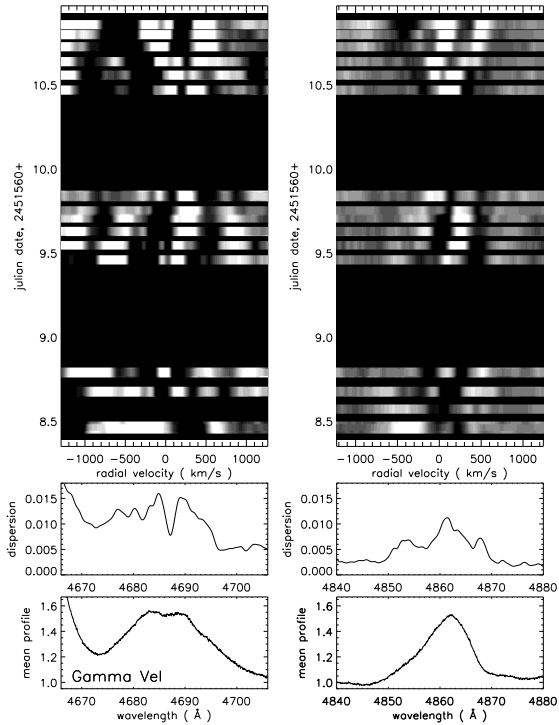}
\caption{Time series of the residuals after subtraction of the mean
  profile, for the Wolf-Rayet star $\gamma^2$ Velorum.}
\end{figure}

\section{Observations}

We observed two Of (one a repeat of previous observations of $\zeta$
Pup) and two late-type, H-rich, weak-line, Galactic WN (super Of-like)
stars intensively during three nights, 2000 January 25/26-27/28, at
the 4m telescope at CTIO. Apart from $\zeta$ Pup, the other three
stars are all located in the Carina Nebula. Parallel observations of
the bright WR star $\gamma^2$Vel (WC8 + O7.5III-V) were also observed
as a check star for known clumps \citep{LEM99}. Basic data on our
five targets are found in Table 1.

We used the bench-mounted \'echelle spectrograph giving an inverse
dispersion of 0.06 \AA\ per pixel and (2.7-pixel) resolving power R
$\sim$30,000 over the $\lambda \lambda$ 4200 - 5700 \AA\ range.  The
slit was 1$\arcsec$ wide. Typical elementary exposures of 10-1000 sec,
depending on the stellar brightness, were combined from the CCD
detector into one spectrum every $\sim$1-2 hours for each star,
yielding net S/N $\approx$ 500 per pixel (no binning) after wavelet
filtering (see below). Wavelengths were calibrated using Th-Ar
spectra. An internal quartz-lamp was used for flat-fielding.

The bright \'echelle B3V flux standard $\eta$ Hya \citep{Hetal92} was
observed in an attempt to calibrate and splice together the \'echelle
orders; however, its standard flux values turned out to be of too low
resolution to be useful, so its use was abandoned. This forced us to
treat each \'echelle order separately. In any case, this was not a
serious impediment, as we chose to focus on the strongest, most
isolated optical emission lines of HeII 4686 and HeII/H$\beta$ 4860 in
each star.  After some experimentation, we obtained final extracted
spectra of good quality by dividing the spectra in order 9 by those
in order 8 for HeII 4686 and order 7 by order 6 for HeII/H$\beta$
4860. In all the stars, orders 8 and 6 were essentially devoid of
emission lines and thus could be used as continuum reference
spectra. This division by neighboring orders allowed us to eliminate
the strong order sensitivity-function, and rectify the spectra without
polluting the target emission lines significantly. 

Nightly mean spectra for each of our 5 targets, are displayed in
Figure 1 for the \'echelle orders covering the HeII 4686 and
HeII/H$\beta$ 4860 atomic lines. All stars show those atomic lines
strongly in emission, with large equivalent widths and very broad
profiles which are typical of Wolf-Rayet and Of stars with strong,
radiation-driven winds. Profiles from the Wolf-Rayet stars are
dominated by the emission component, while profiles from the two Of
stars show a slightly blue-shifted dip which is due to P Cygni type 
absorption from the stellar wind.

\section{Analysis and results}

We first examine whether there are significant global line variations
from night to night in the P Cygni component, as seen before in the
nightly spectra of $\zeta$ Pup \citep{MM81,ELM98}. The nightly means from
each of the three nights are shown in Fig.1 in different
colors. Significant night-to-night variations are observed in the WC
star $\gamma^2$Vel and in the two Of stars $\zeta$Pup and HD
93129A. Variations are also apparent in the HeII 4686 line of the two
WNha stars HD 93131 and HD 93162. The HeII 4686 line in $\zeta$ Pup is
found to display the most pronounced variations, which suggest that
the intense line-profile variations discovered by \citet{ELM98} might
be an exceptional case.

Variations on timescales of hours turn out to be much more interesting
in all the stars. For each spectrum collected during the observing
run, we subtracted the mean profile for the run. Despite all the
calibration steps described in \S2, many instrumental artifacts remain
in the spectra at the $\lesssim$1\% level. While these artifacts have
little effect on the mean spectra, they do show up prominently in the
residuals. Most artifacts are either of low or high frequency in
wavelength/pixel space. We thus filtered all of these out below 0.5
\AA\ and above $\sim$8 \AA\ using wavelet transforms following the
method of \citet{LEM99}.  This is fortunate, given that most of the
intrinsic variations in Wolf-Rayet and Of star lines occur as narrow
emission line features (subpeaks) on a 1-2 \AA\ scale.

\begin{figure*}
\epsscale{1.0}
\plottwo{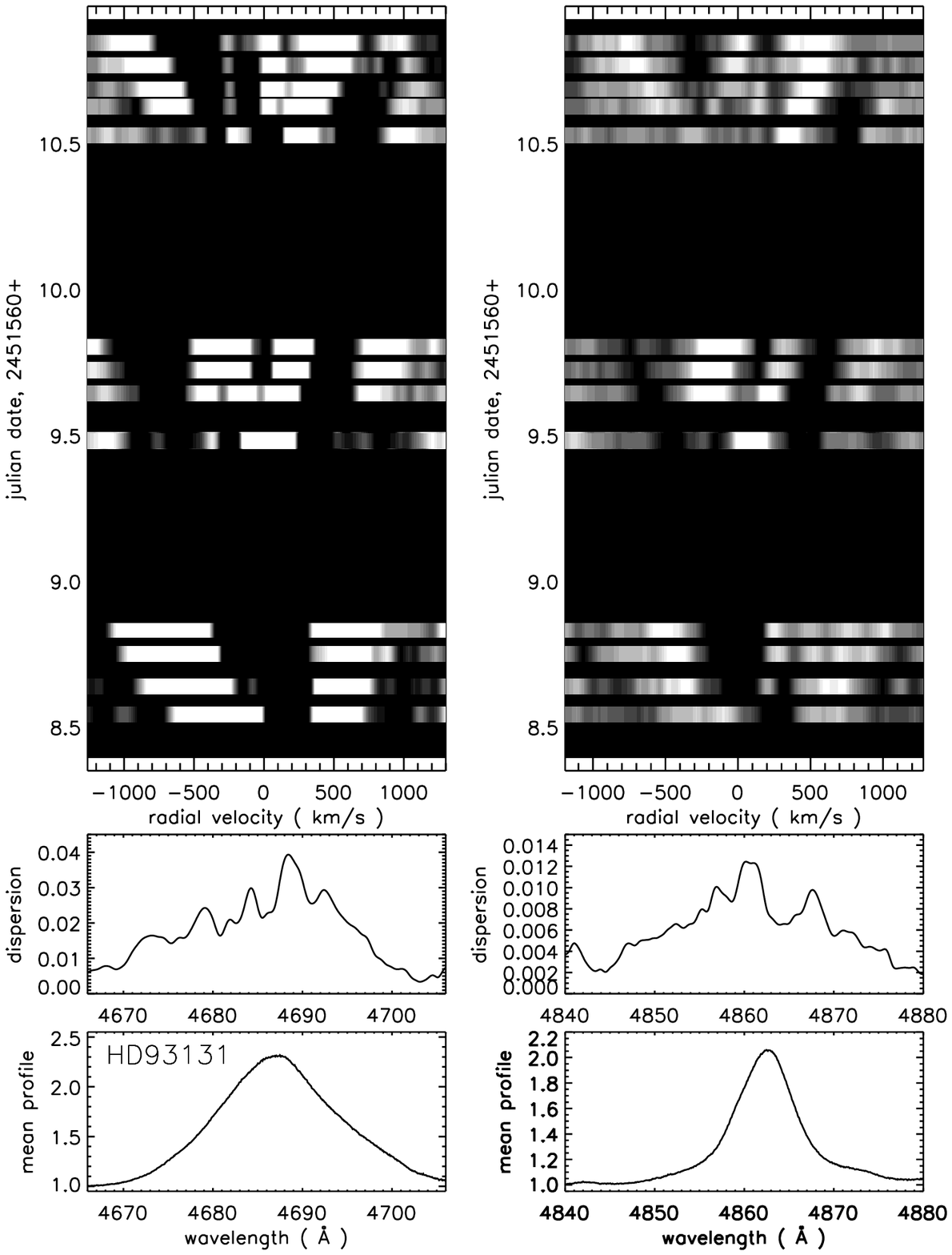}{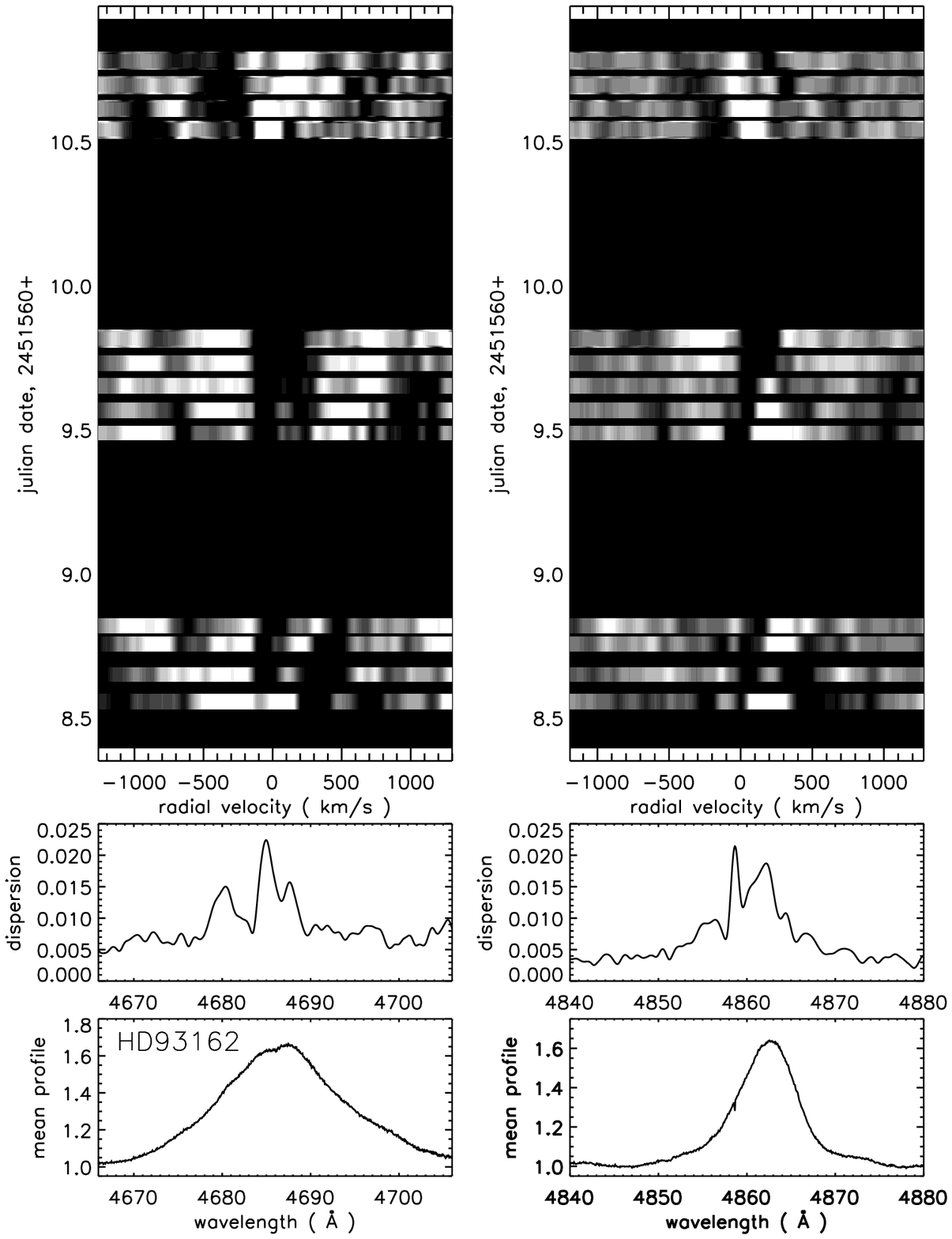}
\caption{Time series of the residuals after subtraction of the mean
  profile, for the H-rich WN stars HD93131 (left) and HD93162 (right).}
\end{figure*}

\begin{figure*}
\epsscale{1.0}
\plottwo{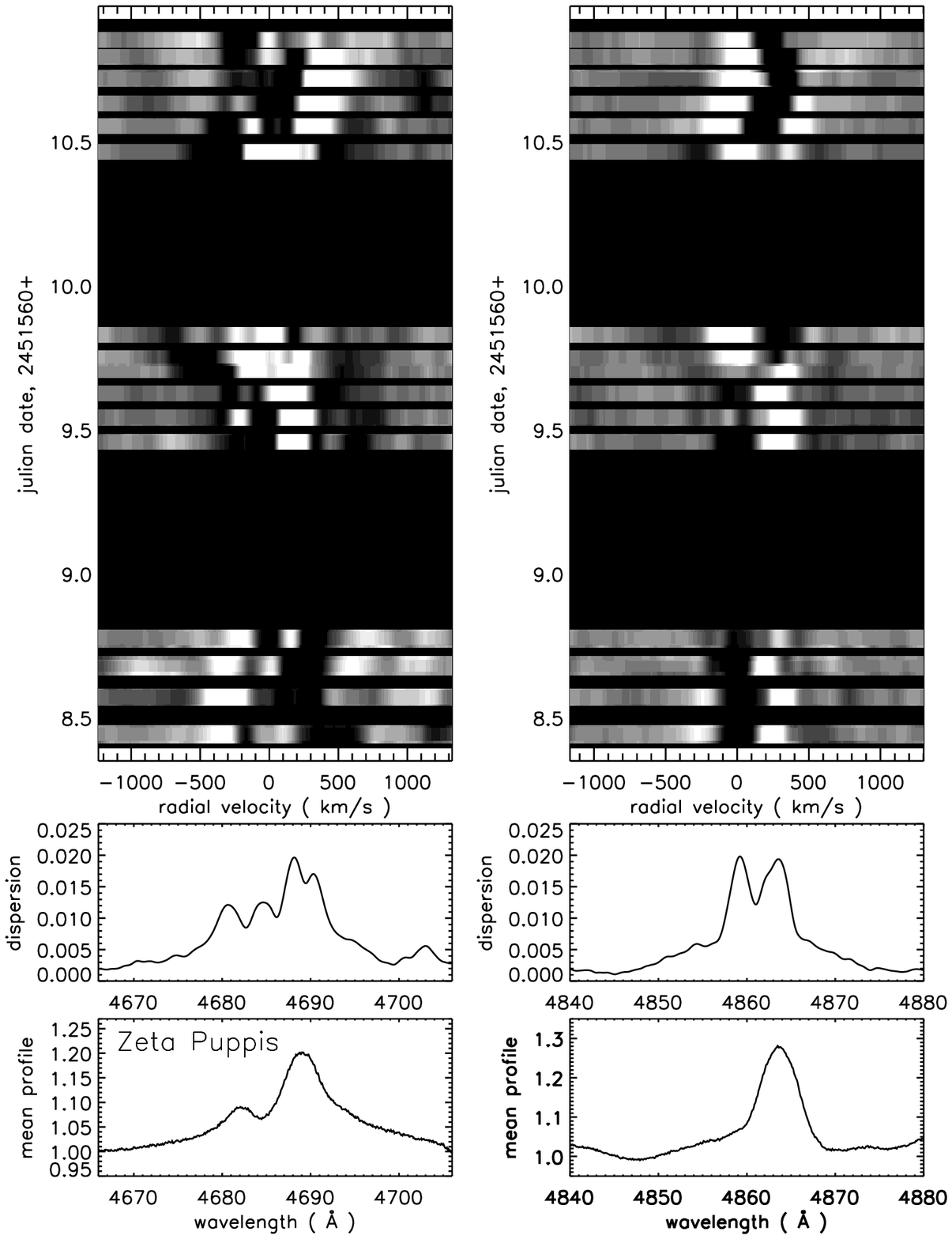}{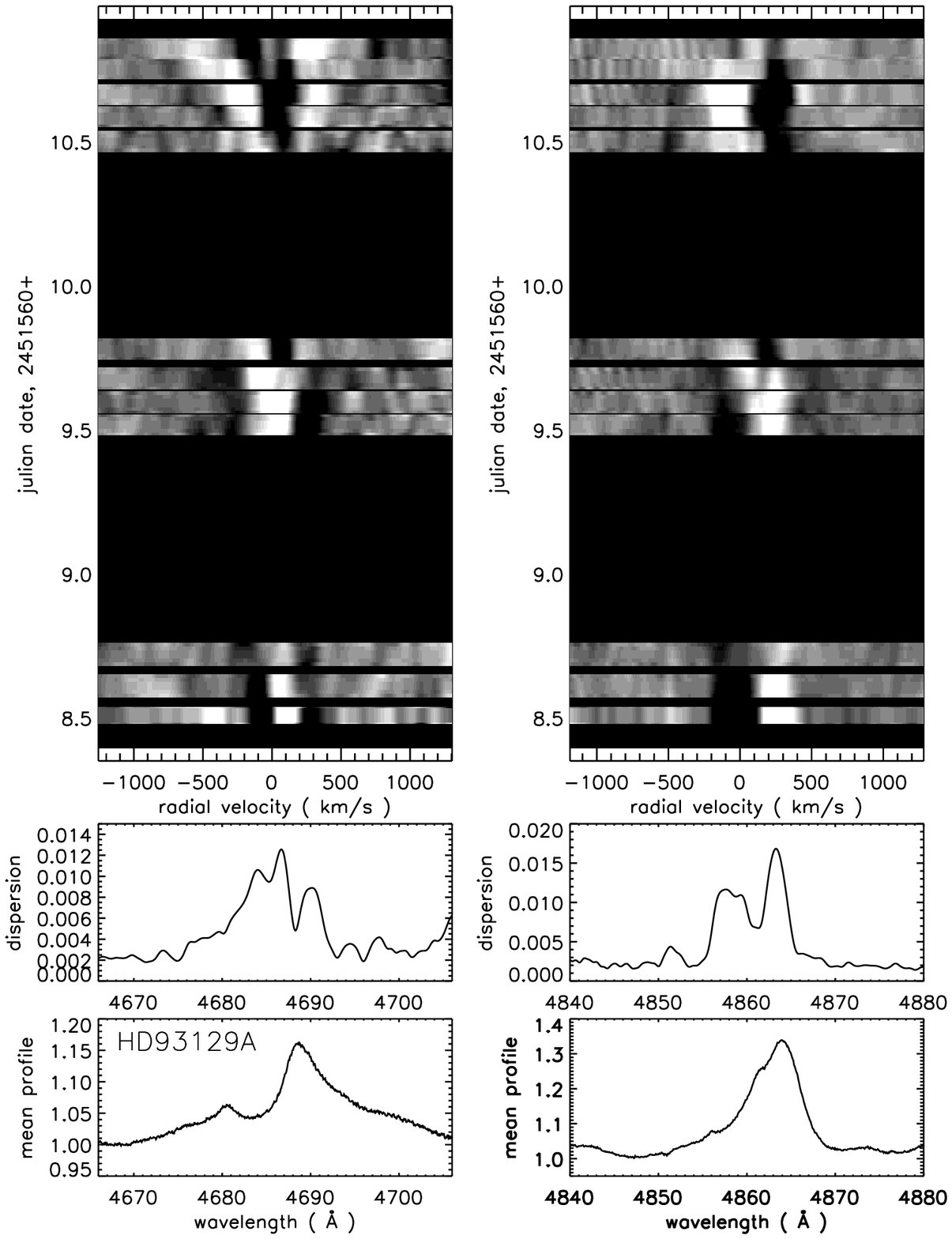}
\caption{Time series of the residuals after subtraction of the mean
  profile, for the Of stars $\zeta$ Puppis (left) and HD93129A (right).}
\end{figure*}

In Figs. 2, 3, and
4, variability in the emission lines of HeII 4686 (left) and
HeII/H$\beta$ 4860 (right) are shown for each of our target stars. The
mean profile for each line is displayed in the bottom panel. The
middle panel plots the dispersion at each pixel for the entire
observing run. The top panel is a two dimensional plot of the residuals
(i.e. the instantaneous deviation from the mean profile) as a function
of time and wavelength. Time runs upward on the vertical axis. The
horizontal axis is labeled in units of radial velocity, calculated
based on the redshift/blueshift of the emission line under
consideration.

Significant variations are observed in all the lines examined. With
variability occurring on timescales of hours, it is clear that at least
some of this variation will average out in the nightly mean. The
night-to-night variations noted in Figure 1 thus underestimate the
variability in the line profiles. Conversely, the mean from all three
nights can be confidently used as a base profile, and the residuals
from this means (as shown in Figs.2, 3, and 4) will provide a good
representation of the variable elements, be they either excess
emission subpeaks or excess absorption troughs.

In all the target stars, both the HeII 4686 and HeII/H$\beta$
4860 lines show variability which is extremely similar to that
previously observed in WR spectra. The same general pattern of
propagation outwards from the line center is apparent in all the stars
for at least one of the nights. These indicate outwards acceleration
in the winds of all these stars. According to the phenomenological
models of \citet{LM99}, emission subpeaks in the line profile do not
necessarily trace the acceleration of single, overdensities (clumps)
in the wind. Rather, the emission subpeaks are likely the combined
signature of numerous clumps, so the outwards motion of subpeak
elements across the line may not be a measure of the true wind
acceleration. In any case, in the one star where the acceleration is
most apparent (HD 93131) subpeak profiles on the blue and red edges
are observed to move by about 400 km/s over the course of one night (8
hours). This suggests an outward wind-acceleration rate on the order
of 14 m s$^{-2}$, which is consistent with acceleration rates
measured in other Wolf-Rayet stars \citep{LM99}.

\begin{figure*}
\epsscale{1.15}
\plottwo{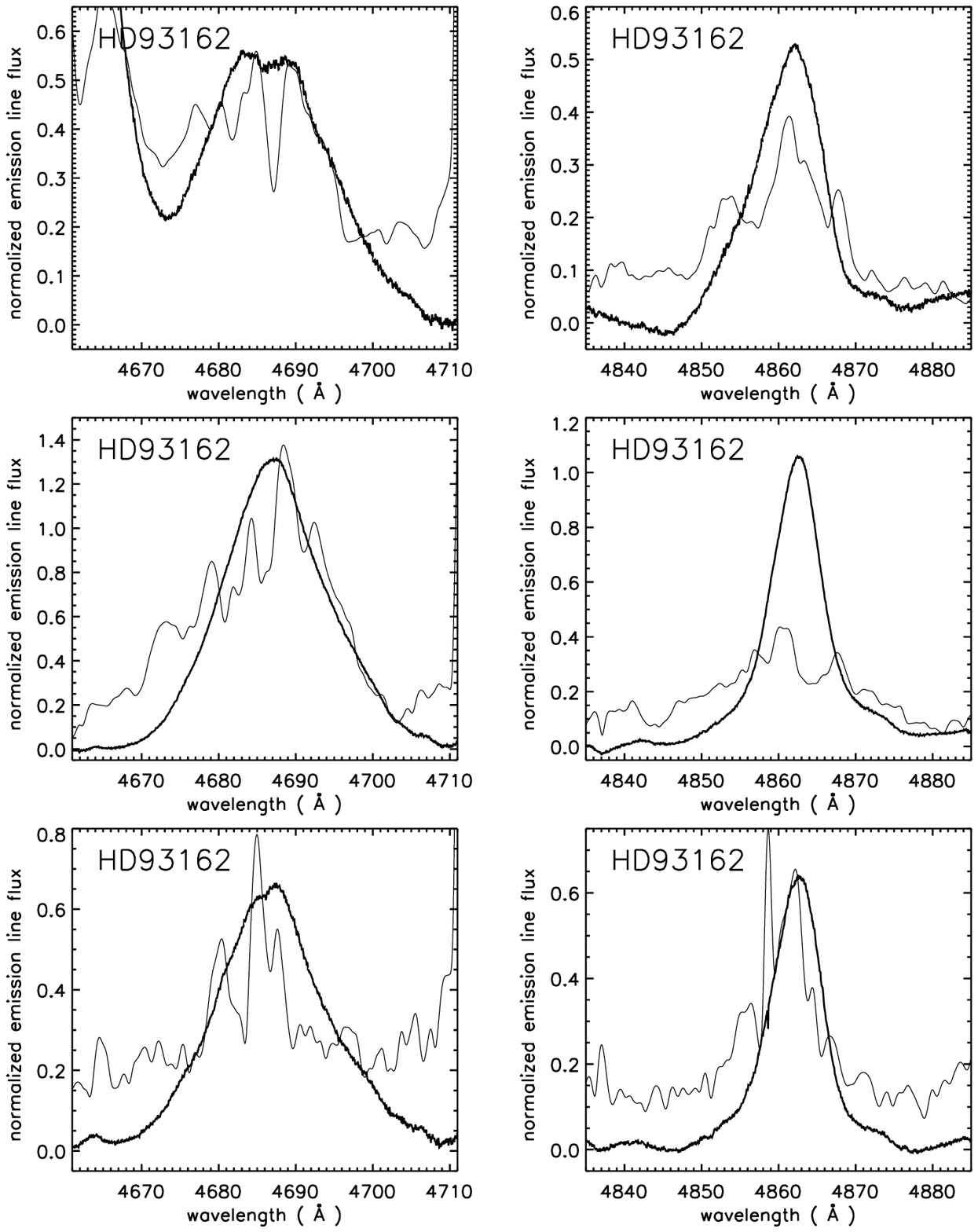}{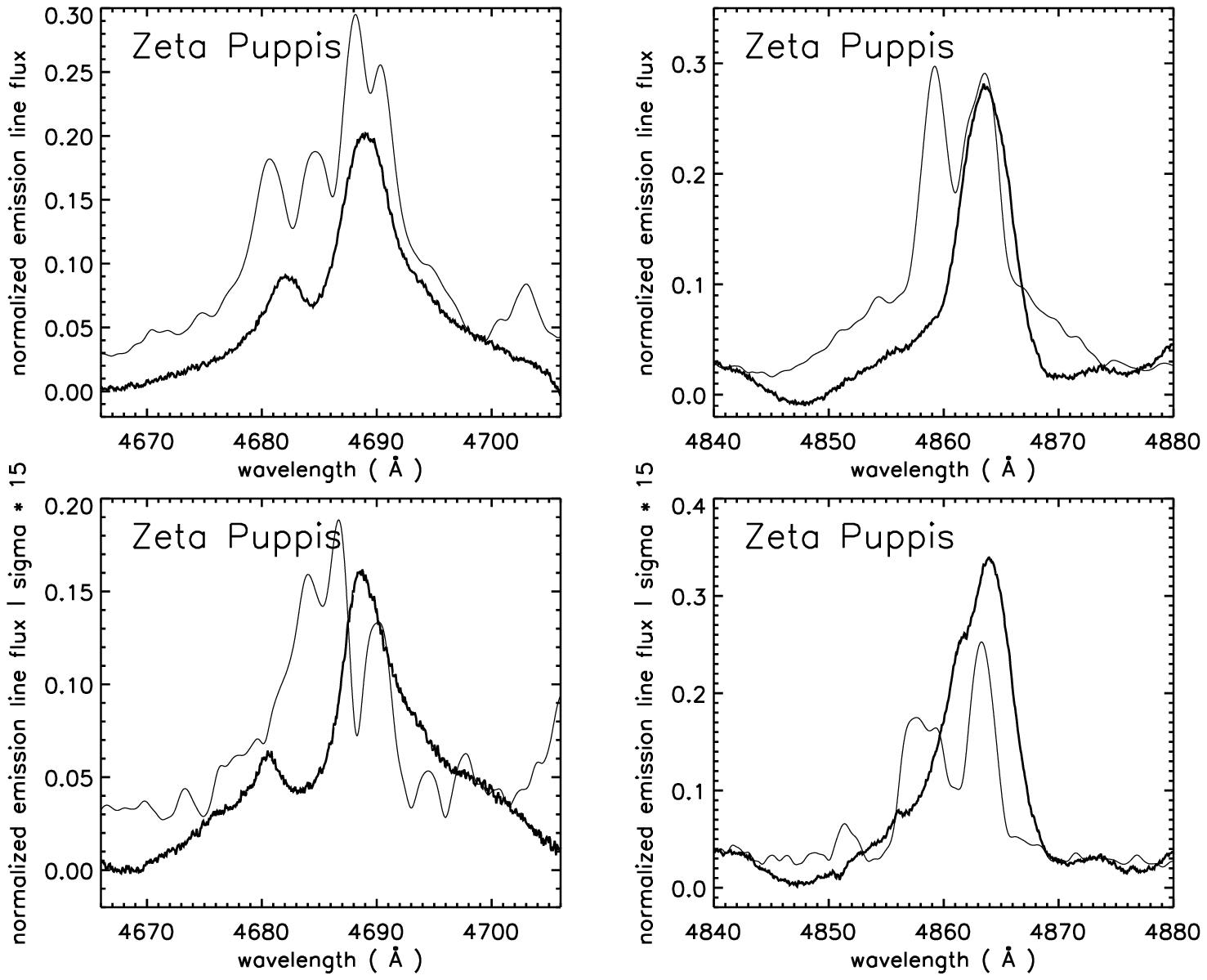}
\caption{
Comparison between the normalized mean profiles of the emission lines
and the observed dispersion in the line profile, as a function of
wavelength. Mean (thick) lines and dispersion (thin) lines are
superposed on each plot, with the amplitude of the dispersion
multiplied by a factor of 35 (hence a dispersion level of 0.35 on the
plot actually corresponds to a dispersion amplitude of 0.01 in
normalized flux units).
}
\end{figure*}

In Figure 5, we condense the variable information into plots of
standard deviation per pixel across each line profile for all spectra
collected over the three nights. Standard deviation profiles are all
multiplied by a factor of 35, and compared with the mean profiles for
each star. On this scale, the deviation profiles and mean
emission-line profiles have similar amplitudes, which suggest that all
lines show variability at the $\sim$3\% level. This level of
variability is on a par with previous data for 9 WR stars
\citep{LM99}, the Of star $\zeta$ Pup \citep{ELM98}, and $\gamma^2$
Vel \citep{LEM99}. The only exception is the He II $\lambda4860$
profile in HD93162, whose variability is below the 1\% level. 
 
\section{Discussion and conclusions}

Our spectroscopic time-series show systematic variability on a
timescale of hours in bright emission-line profiles from hot, massive
stars with moderately strong winds. The emission lines examined
here include lines which are significantly fainter, relative to the
continuum, than the bright lines from Wolf-Rayet stars, where the
phenomenon was first examined \citep{LM99}. We find that the amplitude
of the variations are at the 1\%-3\% level of the line emission. This
suggests that variability is universal in all emission lines, but that
it is only apparent in the strongest of the lines.

While the variability profiles ($\sigma (\lambda)$), after rescaling,
generally follow the underlying line profiles ($I(\lambda)$), they do
not match exactly on close examination. There are two possible causes
for this: (1) the variability is strictly proportional to line
emission, but the profile is distorted by an underlying
absorption component, and/or (2) there exists an additional source of
variability in the line profile, e.g. because of variations in an
underlying absorption component, as in P Cygni absorption edges
\citep{R92}. The effects from distorted line profiles are most
apparent in the lines of the Of stars, which are flanked by a blueward
absorption component, and are strongly asymmetric relative to the rest
wavelength. Despite this, however, we clearly observe (Fig.4) that the
variability profile is roughly symmetric about the rest wavelength,
which confirms our assertion that the variability is associated mainly
with the {\em emission} component.

Overall, it appears that the clumps in the winds that cause the
variable subpeaks across the emission lines, do indeed trace the wind.
This would suggest that, within their turbulent speeds, the clumps
obey essentially the same acceleration and velocity laws as the
average wind itself.  Indeed, the clumps {\it are} the wind!  In fact,
using extremely simple assumptions, \citet{LM99} were able
to simulate strong WR emission lines {\it and their observed
  variability} solely using a superposition of a very large number of
discrete wind-emitting elements, without recourse to any kind of
continuous background wind. This may not be proof of concept, but at
least it is consistent with other astrophysically turbulent media such
as the ISM \citep{S93}. We would thus claim that hot-star
winds are the (supersonic compressible) turbulent result of some kind
of energy-input driving. Whether this occurs predominantly as a process
of cascading dissipation from large to small scales \citep{H94} or of
shock-merging aggregation from small radiative instabilities on
sub-Sobolev \citep{O94} to large scales \citep{F05}, remains to
be demonstrated observationally.

The ratio $\sigma(\lambda) / I(\lambda)$ is observed to vary between
different lines, though only by a factor of a few at most for the same
line in different stellar winds. The more pronounced case here is the
star HD193162 in which HeII 4686 is more variable than HeII/H$\beta$
4860 by about a factor of 3. This could be related to enhanced HeII
4686 emission from the highly turbulent zone of wind-wind collision
between the components in this massive binary system. The variability
levels are however remarkably similar between the various stars
observed here. Since the intensity variations across optical
(recombination) spectral lines vary with the square of the density,
the clumping filling factor, which varies as the square root of this
quantity, must actually show much less dispersion compared to the
observed line-intensity variations, from one wind to another.  From
this, we deduce that most hot-star winds (O, WR, LBV) require nearly
the same reduction factor in mass-loss rate of typically $\sim$3 (2-5
in the extreme) when the mass-loss rates are based on density-squared
mechanisms. This result is in accord with the consensus on the
ubiquity of the clumping phenomenon that came out of the recent
workshop devoted entirely to clumping in hot-star winds \citep{H08}.

Our conjecture is that spectroscopic observations with even higher
signal-to-noise ratio should reveal similar variability patterns in
all emission-line profiles of stars with hot winds. Wind clumping is
the probable source of the variability. The timescale is set by the
time it takes for individual clumps to propagate through the layer in
the wind from which line emission occurs. Different lines will show
different patterns, depending on the height and depth of the layer
where emission occurs. Emission lines formed in overlapping layers
will have similar line-profile variability patterns. From a typical
timescale of $\sim$6 hours and a wind velocity speed of $\sim1000$km
s$^{-1}$, the typical depth of the emission layers is $\sim$2 10$^7$
km, or about 30 R$_{\sun}$, consistent with the idea that the lines
are formed relatively close to the hydrostatic surface. All of these
things should be tested with better future data.

\acknowledgments

{\bf Acknowledgments}

The authors thank the referee, Ken Gayley, for useful comments. AFJM
is grateful for financial assistance from NSERC (Canada). SL is
supported by the U.S. National Science Foundation, through grant
AST-0607757.


\end{document}